\documentclass[letterpaper]{article} 
\usepackage{aaai2026}  
\usepackage{times}  
\usepackage{helvet}  
\usepackage{courier}  
\usepackage{multirow}
\usepackage[hyphens]{url}  
\usepackage{graphicx} 
\usepackage{amsmath} 
\usepackage{natbib}  
\usepackage{caption} 
\usepackage{algorithm}
\usepackage{algorithmic}
\usepackage{subcaption}

\newcommand{\minisection}[1]{\vspace{.06in}\noindent{\textbf{#1}}.}

\usepackage{booktabs}
\usepackage[table]{xcolor}
\usepackage{subfig}
\usepackage{cite}        
\usepackage{tikz}        
\usepackage{etoolbox}    

\nocopyright

\newcommand{\fancycite}[1]{%
  \cite{#1}%
}

\usepackage{newfloat}
\usepackage{listings}
\DeclareCaptionStyle{ruled}{labelfont=normalfont,labelsep=colon,strut=off} 
\floatstyle{ruled}
\newfloat{listing}{tb}{lst}{}
\floatname{listing}{Listing}
\title{Faver: Boosting LLM-based RTL Generation with \\ Function Abstracted Verifiable Middleware}

\author{
    Jianan Mu\equalcontrib \textsuperscript{\rm 1}, Mingyu Shi\equalcontrib \textsuperscript{\rm 2}, Yining Wang \textsuperscript{\rm 3},\\ Tianmeng Yang \textsuperscript{\rm 4}, Bin Sun\textsuperscript{\rm 1}, Xing Hu\textsuperscript{\rm 1}, Jing Ye \textsuperscript{\rm 1}, Huawei Li\textsuperscript{\rm 1}
}
\affiliations{
    \textsuperscript{\rm 1}{the State Key Lab of Processors, Institute of Computing Technology, Chinese Academy of Sciences}
    \textsuperscript{\rm 2}{Nanjing University, School of Integrated Circuits}
    \textsuperscript{\rm 3}{University of Toronto}
    \textsuperscript{\rm 4}{Peking University}

}

\usepackage{bibentry}

\begin{document}

\maketitle

\begin{abstract}

LLM-based RTL generation is an interesting research direction, as it holds the potential to liberate the least automated stage in the current chip design.
However, due to the substantial semantic gap between high-level specifications and RTL, coupled with limited training data, existing models struggle with generation accuracy. 
Drawing on human experience, design with verification helps improving accuracy. However, as the RTL testbench data are even more scarce, it is not friendly for LLMs. Although LLMs excel at higher-level languages like Python/C, they have a huge semantic gap from RTL. 
When implementing the same functionality, Python/C code and hardware code differ significantly in the spatiotemporal granularity, requiring the LLM not only to consider high-level functional semantics but also to ensure the low-level details align with the circuit code. It is not an easy task.
In this paper, we propose a \textbf{f}unction \textbf{a}bstracted \textbf{ver}ifiable middleware (\textbf{Faver}) that streamlines RTL verification in LLM-based workflows. By mixing LLM-friendly code structures with a rule-based template, Faver decouples the details of circuit verification, allowing the LLM to focus on the functionality itself. In our experiments on the SFT model and open-source models, Faver improved the model’s generation accuracy by up to 14\%.

\end{abstract}

\section{Introduction}

RTL design is the process of encoding a functional specification into Register Transfer Level (RTL) code. It remains the least automated and most labor-intensive stage in integrated circuit (IC) chip design. Recently, an increasing number of studies have explored applying large language models (LLMs) to automate RTL design. 
By systematically mining training datasets and employing diverse training strategies, specialized and commercial models have steadily advanced their RTL-generation capabilities~\fancycite{chang2023chipgpt, liu2023chipnemo, pei2024betterv, ho2025verilogcoder, guo2025deepseek}.

\begin{figure}[htbp]
    \center
    \includegraphics[width=0.8\linewidth]{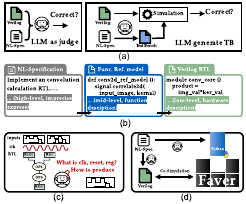}
    \caption{(a) Using the LLM itself as a judge or writing RTL as the testbench both hinder efficient design-with-verification. (b) Motivation: LLM writes high-level code to verify RTL design. (c) Challenge: Traditional high-level code lacks timing-related inputs and variables. (d) Faver: Function abstracted verifiable middleware.}
    \label{fig1_new}
\end{figure}
Given the limited availability of RTL-oriented data, an alternative to merely training models is to adopt a ``design with verification'' approach during inference. This idea mirrors human design experience and has proven effective in improving the correctness of LLM-based software programming~\fancycite{chen2022codet,gou2023critic,wang2023codet5+}. Two straightforward methods for LLM-based RTL design with verification are shown in Figure~\ref{fig1_new}(a): One is to employ the LLM as a judge to directly evaluate whether the design meets the spec, and the other is to have the LLM generate an RTL testbench (encompassing both a reference model and the necessary test stimuli). Unfortunately, our experiments indicate that neither strategy effectively boosts the efficiency of LLM-based RTL generation. Creating an RTL testbench can be just as demanding as writing the RTL design itself, not least because testbench data is even more limited than design data. Moreover, significant challenges persist, including the large semantic gap between natural-language specifications and RTL, along with a lack of robust reasoning tools for validating functionality based solely on specs.

A natural approach, illustrated in Figure~\ref{fig1_new}(b), is to write a function testbench in a high-level language such as Python or C. This offers three advantages:
(1): Validating functionality at a higher level, aligning with human verification practices.
(2): From both semantic-gap and data-abundance perspectives, generating Python/C is preferred by most LLMs.
(3): Leveraging Python–RTL co-simulation frameworks~\fancycite{snyder2004verilator,williams2002icarus} for precise matching rather than relying on the LLM as a judge.
Taken together, these advantages make LLM-based RTL generation look promising. 

However, realizing this “one-stone-three-birds” ambition is far from straightforward. Our further experiments show that simply asking an LLM to write Python code to verify the RTL design has significant misjudgments, thus yielding marginal improvements for RTL generation.
The crux of the problem is that hardware and software approach the same functionality with different spatiotemporal granularities, creating a gap that complicates co-verification. Specifically, as shown in Figure~\ref{fig1_new}(c), there are two challenges.
The first one is about Time-dependent variables. In hardware, inputs are clock-driven bitstreams, and outputs depend not only on the current cycle’s input but also on the internal register state (influenced by prior inputs). However, standard Python/C lacks built-in clock, register, and reset semantics. Although some hardware engineers have created domain-specific languages for high-level verification~\cite{rosser2018cocotb}, such data remains scarce, posing hurdles for LLM-based design and verification.
The second one is about test stimuli. Designing high-quality circuit test stimuli requires simultaneously addressing the high-level temporal semantics and the bit-level precision, which is neither a trivial task.

To this end, we propose Faver, a function abstracted middleware for RTL design with verification. 
It allows the LLM to concentrate on writing high-level semantic code that verifies the circuit. 
Its core method is to encapsulate the differences between the circuit and high-level code into a function-class package.
Overall, we built an effective ``design with verification'' framework for LLM-based RTL generation based on Faver. 
Our experimental results show that Faver improves RTL-generation correctness across different test sets and models.
The contributions of this paper can be summarized as follows:
\begin{itemize}
    \item We introduce Faver, allowing the LLM to write high-level semantic code to verify the circuit and then benefit from the framework of design with verification.

    \item We devise a function-class abstraction template that maps clock- and register-like semantics from hardware design into event-driven Python/C function class, reducing the spatiotemporal gap between hardware and software verification.
    
    \item Experimental results across multiple test sets and LLMs demonstrate that Faver boosts the accuracy of LLM-based RTL generation by up to 14\% on different models.

\end{itemize}

\section{Preliminary}

\subsection{Overview of RTL Design}
Register Transfer Level (RTL) serves as a fundamental abstraction in modern digital hardware design, describing the flow of data and control signals within a circuit across clock cycles. RTL is typically expressed using hardware description languages (HDLs), such as Verilog, which is the most widely adopted in contemporary designs. 

An RTL description consists of two main components: sequential registers and combinational logic. Registers are key elements that capture and store data based on clock signals, thus dictating the sequential behavior of the circuit. Combinational logic, on the other hand, performs arithmetic operations between registers, enabling the execution of more complex tasks within the design.

\subsection{LLM-Based RTL Generation}
Recent advancements in large language models (LLMs) have led to promising results in automating RTL generation. Several foundational models, enhanced by techniques such as supervised fine-tuning (SFT) and reinforcement learning (RL), have demonstrated a certain capability to generate RTL code~\fancycite{liu2023chipnemo,Thakur,VeriGen, guo2025deepseek,RTLCoder}. 
Recently, several works have leveraged a design-with-verification approach to improve model generation, such as Verilogcoder~\fancycite {ho2025verilogcoder} and MAGE~\fancycite{zhao2024mage}. They generate and test, feeding verification results back to the RTL model for notable performance gains. However, these approaches typically rely on manually crafted testbenches, which can be labor-intensive to generate the design.

A few studies now focus on LLM-based RTL verification. For instance, VerilogReader~\fancycite{ma2024verilogreader} generates test vectors to boost coverage and assess design robustness. Another recent work attempts to have the LLM produce Python for verification, but by simply predicting outputs, it fails to deliver accurate cycle-level results. Moreover, the testbench is repeatedly rewritten based on RTL outputs, causing it to converge too closely to the DUT and leading to higher false-positive rates.
More recently, one study attempted to let the LLM generate Python-based testbenches to verify RTL code~\fancycite{qiu2025correctbench}. However, because it did not employ Python-Verilog co-simulation and instead relied on the LLM as a judge to compare the testbench and design outputs, accuracy suffered. Later, some designs are built upon that work and introduced improvements~\fancycite{zhao2025proV}.

\subsection{Human Experience in Design with Verification}
Human engineers often follow a ``design with verification'' flow.
To manage this complexity and improve reliability, established practices include separating design and verification teams to maintain independence, developing high-level reference models and test stimuli early in the process, and conducting co-simulation to validate the Design Under Test (DUT) against the reference model.
Recently, some hardware engineers have begun using Python frameworks with domain-specific languages (DSL) designed for Verilog verification when building testbenches~\fancycite{rosser2018cocotb}. However, these DSLs are scarce in training data, making them less amenable to LLM-based code generation.


\begin{figure*}[tb]
    \center
    \includegraphics[width=1\linewidth]{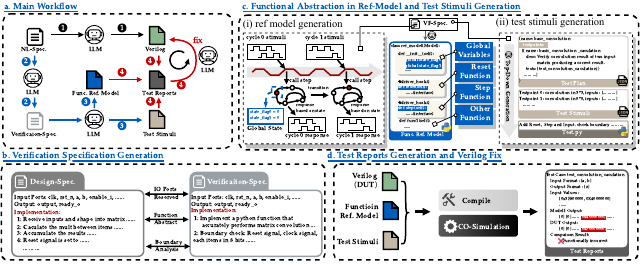}
    \caption{(a) Main workflow of Faver. (b) Verification specification generation. (c) Reference model and test stimuli generation based on class templates that extract time variables.  (d) Test reports generation and Verilog generation recycle.}
    \label{fig2_new}
\end{figure*}

\section{Methodology to Build Faver Framework}\label{sec3}

To approach an efficient and accurate RTL design with verification mechanism, we propose Faver, a system that allows LLM to focus on writing verification in high-level languages such as Python/C. 
As shown in Figure~\ref{fig2_new}(a), Faver has the following key steps:
\begin{itemize}
  \item Step 1: The LLM reads the specification to generate the RTL.
  \item Step 2: The LLM reads the design spec to generate the verification spec (see Figure~\ref{fig2_new}(b)).
  \item Step 3: The LLM uses the verification spec to create the functional reference model and test stimuli(see Figure~\ref{fig2_new}(c)).
  \item Step 4: A Python-Verilog co-simulation is carried out, producing a verification report that is fed back to the LLM for iterative refinement(see Figure~\ref{fig2_new}(d)).
\end{itemize}

We detailly introduce the methodologies of the proposed Faver framework in the following sections, and then conclude with an analysis of Faver’s effectiveness.

\subsection{Verification Specification Generation}
\label{sec_spec}

Instead of having the LLM reproduce the circuit topology in Python verbatim, we want it to write verification code from a higher-level functional perspective. Hence, rather than directly passing the circuit design specification to the LLM for verification code generation, we first translate it into a higher-level verification spec.

\begin{figure}[htbp]
    \center
    \includegraphics[width=0.8\linewidth]{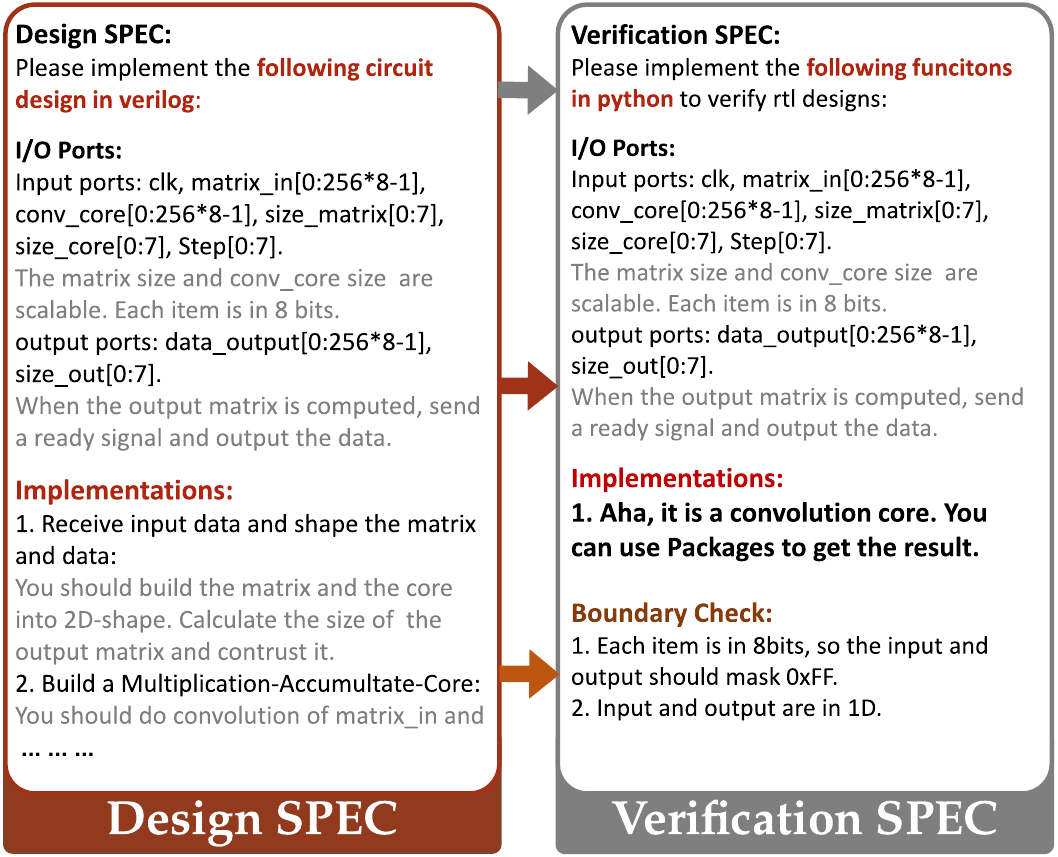}
    \caption{Verification specification generation.}
    \label{fig4_new}
\end{figure}

To ensure that the verification code derived from this higher-level functional description remains consistent with the original design (Figure~\ref{fig2_new}(b)), we propose three key technical components. We will illustrate these using the convolution kernel design example shown in Figure~\ref{fig4_new}.

\minisection{Preserving Input/Output Ports}
We preserve the same I/O port definitions. The input signal clock will be extracted in generating the reference model. 

\minisection{Function Abstraction}
In Verilog, functionality is built up by connecting timing elements in a topology. In contrast, software programming primarily focuses on how inputs are processed to yield outputs. Therefore, based on the design specification, the LLM is required to summarize the corresponding functionality.

As shown in Figure~\ref{fig4_new}, the RTL design spec for the convolution kernel details how it realizes the convolution operation through the interconnection of multipliers and adders. By reading this specification, the LLM can conclude: ``\textbf{Aha, this is a convolution function}''. From then, it can write concise Python code to implement the functionality.

\begin{figure*}[tb]
    \center
    \includegraphics[width=1.0\linewidth]{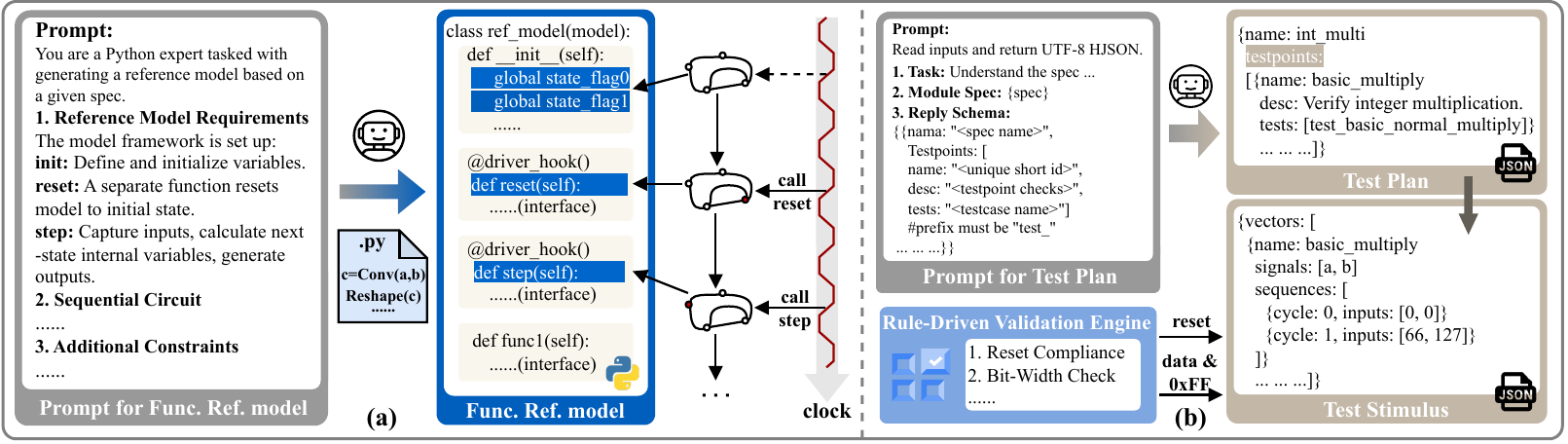}
    \caption{(a) Reference model generation in Faver. (b) Test stimuli generation in Faver.}
    \label{fig5_new}
\end{figure*}

\minisection{Boundary Analysis}
Differences in data specification and formatting between the RTL code and Python code can introduce boundary-condition errors. To avoid such issues, we instruct the LLM to analyze the RTL’s boundary conditions and enumerate them in the verification spec.

For instance, Figure~\ref{fig4_new} shows that in the design of the convolution kernel, each data element has a bitwidth constraint of at most 8 bits, which must be taken into account when generating test inputs. Additionally, Python allows for ample flexibility with high-dimensional tensors, whereas Verilog syntax does not. It only supports fixed-length one-dimensional arrays, so extra care must be taken when matching I/O ports between Python and Verilog.

\subsection{Reference Model Generation Based on Class Templates with Time-Variable Extraction}

In this step, we aim to have the LLM directly generate Python functions that can co-simulate with Verilog that typically features a clock input along with associated operations (e.g., register variables and reset logic). Nevertheless, native Python lacks these constructs and  specialized DSLs to embed these concepts are often scarce and often unfriendly to LLMs. To address these challenges, we equipped Faver with the following designs:

\textbf{Class Template for Clock-Variable Extraction.}  
As illustrated in Figure~\ref{fig2_new}(c)-i, our core mechanism is a class template that includes:
\begin{itemize}

  \item Mapping Register into Global Variables: Register variables are mapped to global (class-level) variables and initialized in the constructor, thereby simulating memory-like, timing-dependent behavior. 
  \item Unified Function Interface: An external \texttt{step} function accepts inputs, performs the design’s functionality, and produces outputs. 
  \item Reset Function: A dedicated \texttt{reset} function can be called externally to re-initialize outputs, global variables, and other stateful components. 
  \item Additional Functions: Various functions remain as stubs for the LLM to fill in, dividing the design into modules accessed uniformly through the \texttt{step} interface.
\end{itemize}

\textbf{Clock as an Event.}  
As shown in Figure~\ref{fig2_new}(c)-i, we treat the clock input as an “event,” thus avoiding input the clock signal into Python codes. Specifically, each rising edge of the clock corresponds to a “\texttt{call step}” action along with an input vector, resulting in a sequence of \texttt{call step} events. Within the class, functionality is invoked through \texttt{step}, which acquires the inputs and returns the outputs.

\textbf{Filling the Class Template.}  
After setting up the class template, we let the LLM fill it with the required logic.

Figure~\ref{fig5_new}(a) demonstrates this process using a convolutional kernel design. 

\begin{itemize}

  \item 1. In Figure~\ref{fig5_new}(a), Faver provides a class template that includes \texttt{global variable declarations} in the init function, a \texttt{reset} function, the \texttt{step} function to catch up with the sequential input series of RTL design, and internal function stubs.
  
  \item 2. As shown in Figure~\ref{fig5_new}(a), together with the verification spec and appropriate prompts, the LLM writes the convolution function and embeds it within the \texttt{step} function while also creating a \texttt{reshape} function for boundary-condition handling. Both are added to the \texttt{step} flow.  
  
  \item 3. Ultimately, the LLM produces a Python-based convolution reference model that seamlessly co-simulates with Verilog.
\end{itemize}

\subsection{Hierarchical Test Stimuli Generation via LLM–Rule Collaboration}

The fundamental challenge in generating test stimuli is to balance high-level semantic coverage and functional relevance with low-level data-stream accuracy. At the high semantic level, test stimuli must exhibit meaningful temporal functions and provide sufficient coverage of the design’s functional space. This is an area where LLMs excel. However, at the fine-grained data-stream level, ensuring the design is reset correctly and handling boundary conditions accurately is difficult for LLMs.

As illustrated in Figure~\ref{fig2_new}(c)-ii, the core idea in this step is to leverage the respective strengths of the LLM and a rule-based generator:

\begin{itemize}

  \item High-Level Planning: Use the LLM to devise a test plan, detailing the functionalities each test case targets. This ensures comprehensive coverage of the functional space. 
  
  \item Temporal Data Generation: Instruct the LLM to produce time-series data for each test case, ensuring that the sequential inputs have strong functional relevance. 
  
  \item Rule-Based Refinement of Stimuli: At the low-level data stream, use specialized rule-based code to refine the stimuli. Specifically:  
   1. Insert a rule-based reset function to ensure the circuit enters a valid operational state.  
   2. Perform boundary checks and fixes on the data stream.
\end{itemize}

As shown in Figure~\ref{fig5_new}(b), for the case of verifying the convolution circuit, firstly, we instruct the LLM to generate test cases covering various convolution specifications at a higher level to ensure broad functional coverage. 
Then, within each test case, LLM provides time-stepped data streams that validate the expected functionality.  
Subsequently, the rule-based code extracts the reset mode (synchronous or asynchronous, active-high or active-low) from the specification and appends the corresponding reset signals to each test case. It also applies boundary checks and fixes to the data stream based on the bitwidth and format constraints found in the specification.  
After that, we obtain a high-quality and accurate set of test stimuli.

\subsection{Co-Simulation, Reporting, and Iterations}

As shown in Figure~\ref{fig2_new}(d), we employ co-simulation to identify any discrepancies between the design and the testbench. If the results match, the verification is passed and the design is approved.

In the event of mismatches, we perform a character-based comparison of the waveforms, classifying whether the error is strictly functional, a timing mismatch, or a boundary-condition issue. We parse the waveforms into data sequences, combine the analysis findings into a single report, and feed it back to the LLM for iterative refinement.

To avoid an infinite loop, we set an iteration threshold. Specifically, once verification is deemed successful, the RTL is taken as the final output. Otherwise, the process repeats. If the number of iteration cycles reaches the threshold, we randomly pick one of the previously generated results to output.

\begin{figure}[htbp]
    \center
    \includegraphics[width=0.90\linewidth]{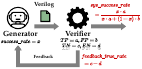}
    \caption{Model of Faver.}
    \label{fig3}
\end{figure}

\section{Analysis of Faver}

\minisection{Model of Faver}
As shown in Figure~\ref{fig3}, Faver can be seen as a system consisting of a generator and a verifier.
The system runs the following rules:
\begin{enumerate}
  \item Output the design that passes verification.
  \item If not passed, give feedback signal, regenerate the design.
  \item After $N$ consecutive failures, sample one design.
\end{enumerate}

Suppose the generator’s success rate is $x$, and the fail rate is $1 - x$ and then the verifier has a true positive rate $\text{TP} = a$, false positive rate $\text{FP} = b$, true negative rate $\text{TN} = c$, and false negative rate $\text{FN} = d$. We can compute the system success rate and the rate of whether the feedback is true ({Feedback True Rate}) as follows:

\minisection{System Success Rate}
The system’s success rate, $\text{sys\_success\_rate}$, can be computed as
\[
\text{sys\_success\_rate} = \frac{x \cdot a}{x \cdot a + (1 - x) \cdot b}.
\]
We have $\text{sys\_success\_rate} > x$ if and only if $a > b$ for $x>0$. 
Therefore, if the checker’s $\text{FP}$ is lower than $\text{TP}$, the system gains a net benefit.

\minisection{Feedback True Rate}
The rate of the feedback is true can be computed as
\[
\text{feedback\_true\_rate} = c - d.
\]
Thus, if the verifier’s $\text{TN}$ surpasses $\text{FN}$, the verifier’s feedback is valuable to the system from the verifier.

\section{Implementation and Experimental Results}

\begin{table*}[htbp]
\centering
\caption{Comparison of Faver against various baselines}
\label{tab:models}
\resizebox{0.8\textwidth}{!}{%
\begin{tabular}{clcccc}
\midrule
    \multirow{2}{*}{Type} & 
    \multirow{2}{*}{Mode} & 
    \multicolumn{2}{c}{VerilogEval} & 
    \multicolumn{2}{c}{RTLLM} \\
 & & pass@1 & pass@5 & pass@1 & pass@5 \\
\midrule
\multirow{5}{*}{Foundation} 
        & DeepSeek-R1-0528 & 0.83 & 0.88 & 0.74 & 0.77 \\
        & Kimi K2 & 0.74 & 0.79 & 0.67 & 0.77 \\
        & GPT-4O & 0.60 & 0.71 & 0.42 & 0.66 \\
        & QWQ-32B & 0.64 & 0.78 & 0.51 & 0.71 \\
        & Qwen2.5-Coder-32B-Instruct & 0.48 & 0.58 & 0.48 & 0.68 \\
\midrule
\multirow{5}{*}{SFT Models}
        & CodeV-CL-7B & 0.45 & 0.60 & 0.40 & 0.62 \\
        & CodeV-DS-6.7B & 0.53 & 0.63 & 0.42 & 0.55 \\
        & CodeV-CQ-7B & 0.53 & 0.65 & 0.37 & 0.55 \\
        & RTLCoder-Mistral-7B & 0.37 & 0.46 & - & 0.48 \\
        & CraftRTL-SC2-15B & 0.68 & 0.72 &0.49 & 0.66 \\
        & Qwen2.5-7B-SFT & - & - & 0.40 & 0.49 \\
\midrule
\midrule
Type & Model & {sys\_sel\_pass@1} & {sys\_inner\_pass@5} & {sys\_sel\_pass@1} & {sys\_inner\_pass@5} \\
\midrule
\multirow{2}{*}{Design with Verification}
        & DeepSeek-R1-0528+Python Verify    & 0.83  & 0.86  & 0.74  & 0.80   \\
        & Kimi K2+Python Verify     & 0.74  & 0.81  & 0.69  & 0.77   \\

\midrule
\rowcolor[rgb]{0.925,0.925,0.925}
        & Faver-DeepSeek-R1-0528    & 0.85 & 0.90   & 0.83   & 0.89   \\
\rowcolor[rgb]{0.925,0.925,0.925}
\multirow{-1}{*}{Ours} 
        & Faver-Kimi K2     & 0.81 & 0.83   & 0.74   & 0.83   \\
\rowcolor[rgb]{0.925,0.925,0.925}
        & Faver-Qwen2.5-7B-SFT  & - & - & 0.51 & 0.63 \\
\midrule
\end{tabular}
}
\end{table*}

We implemented Faver using Toffee~\fancycite{toffee}, a Python-Verilog co-simulation platform based on Verilator and Picker~\fancycite{picker}. Our automated invocation framework, model templates, and prompt generation are all written in Python. 
The threshold for continuous failures before output is set to 5.

We conducted experiments using three different models: Qwen2.5-7B-SFT, DeepSeek-R1-0528~\fancycite{guo2025deepseek}, and Kimi~K2~\fancycite{team2025kimi}. Qwen2.5-7B-SFT was trained on our self-collected data. We conducted SFT on a curated dataset consisting of over 5000 natural language descriptions paired with corresponding RTL code implementations. We adopted the Hugging Face Transformers library as the training framework and utilized LoRA method~\fancycite{huLoRALowRankAdaptation2021} to reduce computational overhead while maintaining performance.

For our experiments, we utilized two Verilog code generation benchmarks from the academic community: RTLLM~\fancycite{lu2024rtllm} and Verilog-Eval~\fancycite{liu2023verilogeval}. Our evaluation metric is the \textit{pass rate}, which indicates that an RTL design’s syntax and functionality both pass the respective checks. More specifically, we report first pass rate (~\emph{Pass@1}) and five pass rate (\emph{Pass@5}).
Furthermore, for the \emph{design with verification} framework, we define two specialized metrics: \textbf{sys\_sel\_pass@1}: The pass rate of a single design output by the \emph{design with verification} system after up to $N$ generations (where $N=5$ in our experiments). Once the system decides on one design for output, if that design passes, we count it as a success. \textbf{sys\_inner\_pass@5}: Across five internal iterations (collectively generating five candidate designs), if any one of these designs passes, we count it as a success.

\subsection{Experimental Results and Comparisons}

Table~\ref{tab:models} presents our experimental results and compares them with alternative methods. To provide a thorough evaluation, we reproduced a verification design directly writing Python-based testbenches~\cite{qiu2025correctbench} without using Faver and integrated it into the design with verification system for comparison (DeepSeek-R1-0528+Python Verify and Kimi K2+Python Verify).
We also tested several baseline models (including DeepSeek-R1-0528~\fancycite{guo2025deepseek}, Kimi~K2~\fancycite{team2025kimi}, GPT-4o~\fancycite{hurst2024gpt}, QWQ-32B~\fancycite{Qwen_QwQ_32B}), along with various SFT-based models specifically tailored for Verilog generation, such as the CodeV~\fancycite{zhao2024codev} series, RTLCoder~\fancycite{liu2024rtlcoder}, the CraftRTL~\fancycite{liu2024craftrtl} series and Qwen2.5-7B-SFT. Among these, Qwen2.5-7B-SFT is the model that we trained ourselves. Note that during the SFT process, a portion of the training data includes samples approximating those in the VerilogEval dataset, posing a risk of data leakage. Therefore, we exclude this benchmark from evaluation for Qwen2.5-7B-SFT (indicated as ``–'' in Table~\ref{tab:models}).
It should be noted that in Table~\ref{tab:models}, the models listed under the \emph{Foundation} and \emph{SFT Models} categories do not use the \emph{design with verification} system; instead, they simply generate RTL designs directly without verification-driven refinement.

At the bottom of Table~\ref{tab:models}, we report the effect of applying Faver to different models. The results demonstrate that, whether on the VerilogEval or RTLLM test sets, Faver consistently improves both \emph{sys\_inner\_pass@5} and \emph{sys\_sel\_pass@1} across all tested models.
For example, DeepSeek-R1-0528 achieves a \emph{pass@1} of 74\% and a \emph{pass@5} of 77\% on the RTLLM dataset, whereas Faver attains a \emph{sys\_sel\_pass@1} of 83\% and a \emph{sys\_inner\_pass@5} of 89\%. Notably, the \emph{sys\_sel\_pass@1} of Faver is \emph{12\% higher than the original pass@5}. 
For Qwen2.5-7B-SFT, Faver's \emph{sys\_inner\_pass@5} is 14\% higher than that of Qwen2.5-7B-SFT.
Overall, Faver introduces about a 10\% improvement on pass rate to the foundation models.

Meanwhile, asking LLM to write a python-based verification \emph{without} Faver fails to provide consistent improvements. For instance, ``DeepSeek-R1-0528 + Python Verify’’ exhibits highly variable gains, often having only a limited positive impact relative to DeepSeek-R1-0528 alone. These observations collectively indicate that Faver not only significantly boosts the correctness rates of LLM-generated RTL designs but also achieves an outcome that direct Python-based verification fails to deliver.

\begin{figure}[tb]
    \center
    \includegraphics[width=1.0\linewidth]{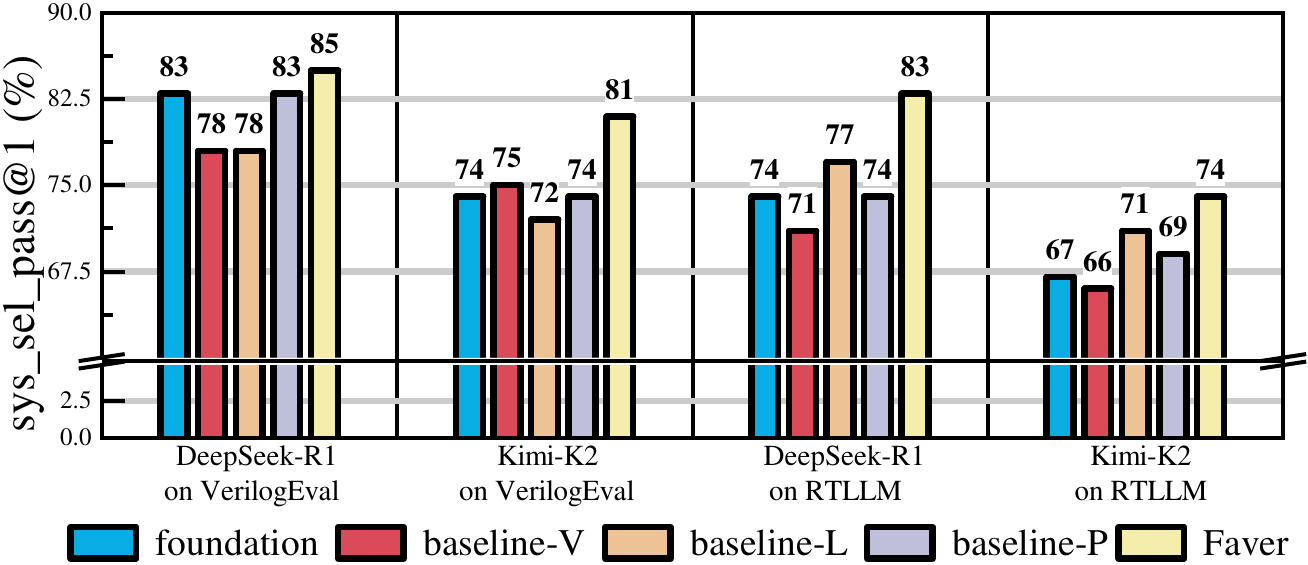}
    \caption{ Analysis on function abstract verification.}
    \label{fig6_new}
\end{figure}

\begin{figure*}[tb!]
    \centering
    \begin{subfigure}[b]{0.44\linewidth}
        \centering
        \includegraphics[width=\linewidth]{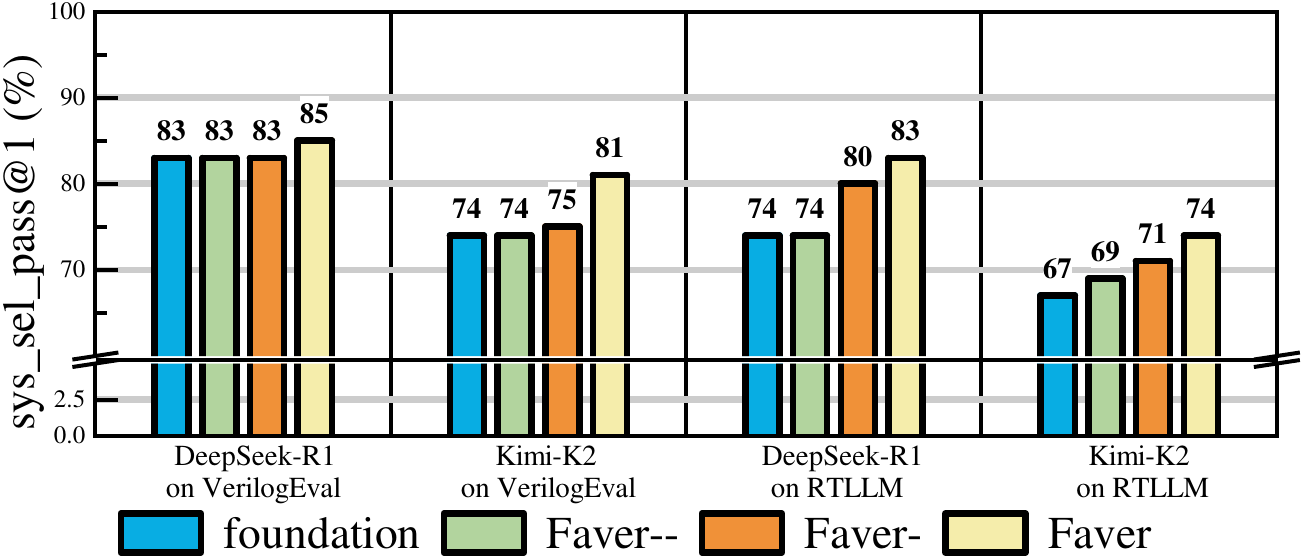}
        \caption{sys\_sel\_pass@1 result}
        \label{fig:exp_1_at}
    \end{subfigure}
    \hspace{1.5em}
    \begin{subfigure}[b]{0.44\linewidth}
        \centering
        \includegraphics[width=\linewidth]{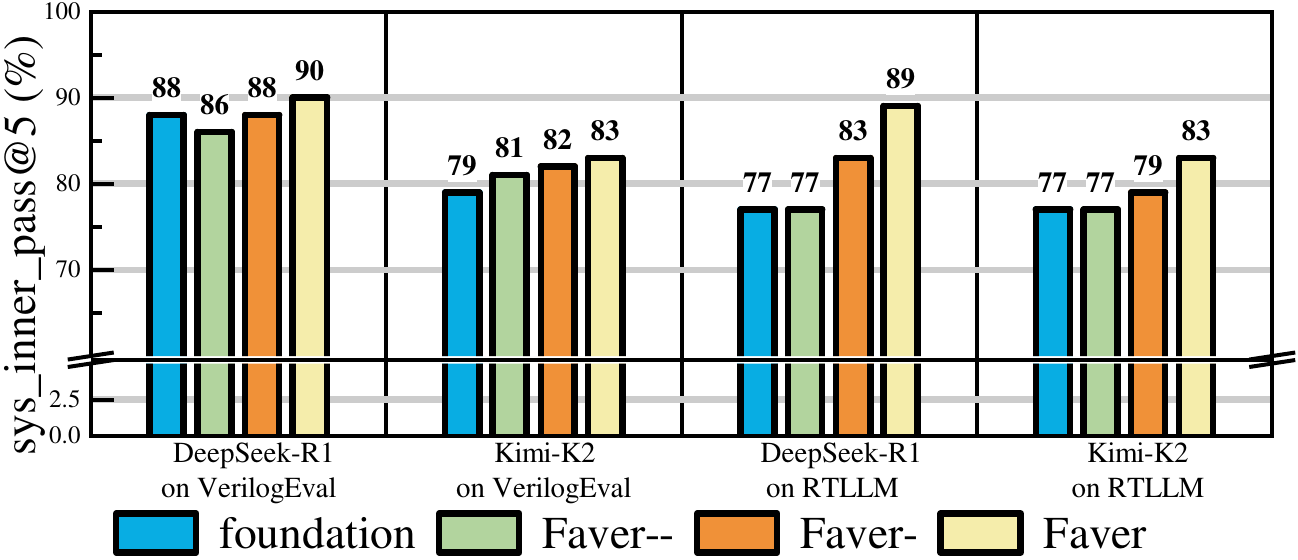}
        \caption{sys\_inner\_pass@5 result}
        \label{fig:exp_1_power}
    \end{subfigure}
    \caption{Ablation study for Faver.}
    \label{fig:exp_ablation}
\end{figure*}

\minisection{Further Analysis on Function Abstract Verification}

We further provide an experimental comparison to highlight the impact of our functional abstraction verification middleware. Figure~\ref{fig6_new} illustrates the \emph{sys\_sel\_pass@1} performance on the RTLLM and Eval datasets when applying the DeepSeek-R1-0528 and Kimi K2 models combined with different approaches. In this setup:
\begin{itemize}

    \item \textbf{foundation}: The base model (no verification).
    \item \textbf{baseline-V}: LLM generate Verilog testbenches for verification.
    \item \textbf{baseline-L}: LLM serves as judge.
    \item \textbf{baseline-P}: LLM generates Python testbenches for verification without using Faver.
\end{itemize}

The results indicate that Faver \emph{consistently outperforms} all compared methods on every test configuration. In contrast, \textbf{baseline V}, \textbf{L}, and \textbf{P} show only limited and somewhat inconsistent improvements over the \emph{foundation} model, underscoring the effectiveness of functional abstraction verification in Faver.

\subsection{Ablation Study}

We further conduct ablation study experiments using two models, DeepSeek-R1-0528 and Kimi K2, evaluated on both the RTLLM and VerilogEval benchmarks. As illustrated in Figure~\ref{fig:exp_ablation}, we compare the following configurations:

\begin{itemize}

    \item \textbf{foundation}: The base model (no verification).
    \item \textbf{Faver}: Entire Faver.
    \item \textbf{Faver-}: The system adopts Faver’s reference model generation framework without using the test stimuli generation framework.
    \item \textbf{Faver}\textbf{-}\textbf{-}: The system employs a direct Python refmodel and stimuli but does \emph{not} leverage Faver.
    
\end{itemize}

Figures~\ref{fig:exp_ablation}\,(a) and~\ref{fig:exp_ablation}\,(b) plot the \emph{sys\_sel\_pass@1} and \emph{sys\_inner\_pass@5} results, respectively. As seen in Figure~\ref{fig:exp_ablation}, for both DeepSeek-R1-0528 and Kimi K2:
Transitioning from foundation to Faver-- (i.e., a simple Python-based testbench without Faver) yields almost no improvements.
Leveraging Faver-, which incorporates Faver’s refmodel generation framework, shows an average performance gain of 2.75\% on both VerilogEval and RTLLM datasets.
Finally, applying the full Faver approach, which also includes Faver’s hierarchical test stimuli generator provides up to a 12\% boost in performance.

These results underline that both the ref model generation and hierarchical test stimuli components of Faver notably increase verification accuracy, thereby leading to more accurate RTL designs overall.

\begin{figure}[htbp]
    \center
    \includegraphics[width=0.8\linewidth]{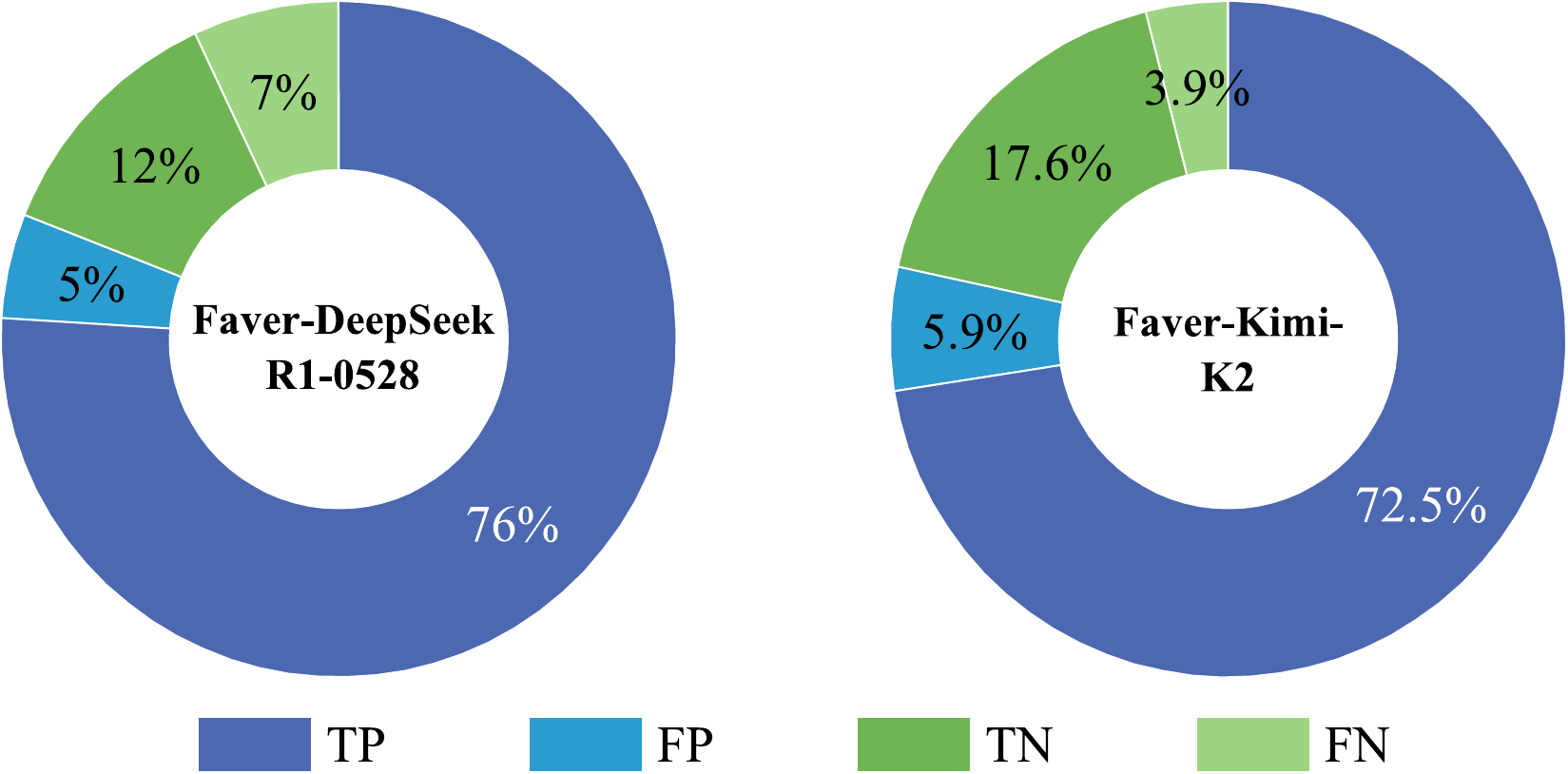} \label{fig:cycle00}
    \caption{TP, FP, TN, FN of the verifier in Faver on VerilogEval.}
    \label{fig_exp_TPFP}
\end{figure}

\subsection{Analysis on success rate of generation and verification }

\begin{figure}[htbp]
    \center
    \includegraphics[width=0.9\linewidth]{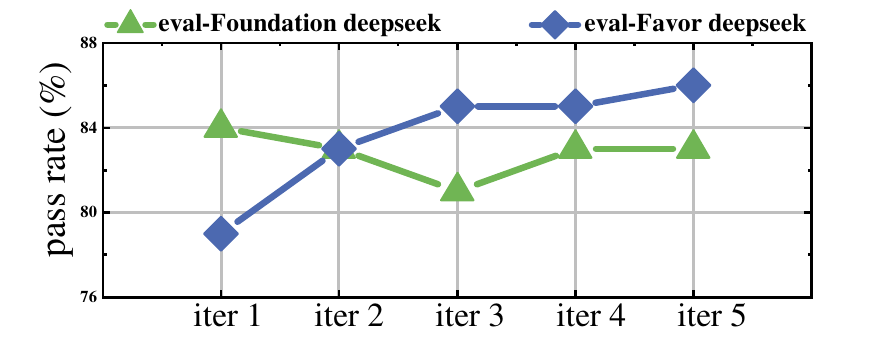}
    \caption{The curve of RTL generation accuracy within Faver over five iterations (data are collected on VerilogEval).}
    \label{fig_exp_rate}
\end{figure}

We further analyze how Faver’s verification accuracy and feedback signals affect the model’s correctness. In Figure~\ref{fig_exp_TPFP}, we present the evaluation results of Faver-DeepSeek-R1-0528 and Faver-Kimi~K2 on the VerilogEval benchmark. The data show that, for both models, Faver consistently yields significantly higher \emph{true positives} compared to \emph{false positives}, as well as higher \emph{true negatives} compared to \emph{false negatives}.

Next, in Figure~\ref{fig_exp_rate}, we compare Faver-DeepSeek-R1-0528 with the original DeepSeek-R1-0528 in terms of the code correctness rate within each iteration of the verification loop on the VerilogEval dataset. We observe that DeepSeek-R1-0528’s accuracy fluctuates in a rather random pattern, whereas the feedback signals provided by Faver-DeepSeek-R1-0528 drive a steady improvement in code correctness.

As illustrated by Figures~\ref{fig_exp_TPFP} and~\ref{fig_exp_rate}, Faver boosts the overall RTL generation accuracy and substantiates our analytical insights regarding the contribution of accurate verification feedback in improving generation quality.

\section{Conclusion}
In this work, we introduced Faver, a functional abstraction middleware that bridges the gap between software-level verification practices and the time- and state-dependent nature of hardware design. 
We devise a function-class abstraction template that maps clock- and register-like semantics from hardware design into event-driven Python/C functions, reducing the spatiotemporal gap between hardware and software verification.
Extensive experiments show that Faver improves the correctness of RTL design by about 5-14\% across various test sets and models. Importantly, this work demonstrates that precise function-level verification feedback is crucial for refining LLM outputs in hardware design contexts. 
We hope Faver can ``do a favor'' for LLM-based RTL generation in real-world applications in the future.



\newpage
\bibliography{aaai2026}



\end{document}